\documentclass[12pt]{iopart}
\usepackage{graphicx}

\newcommand\beq{\begin{equation}} \newcommand\eeq{\end{equation}}
\newcommand\bea{\begin{eqnarray}} \newcommand\eea{\end{eqnarray}}
\renewcommand\aa{{\cal A}} 

\begin{document}


\title[Semiclassical corrections to the interaction energy of a
hard-sphere Boltzmann gas]{Semiclassical corrections to the
  interaction energy of a hard-sphere Boltzmann gas} 

\author{R K Bhaduri\dag, W van Dijk\ddag\dag, and M K Srivastava\S}

\address{\dag\ Department of Physics and Astronomy, McMaster
  University, Hamilton, Canada \ L8S~4M1} 
\address{\ddag\ Physics
  Department, Redeemer University College, Ancaster, Canada \ L9K~1J4}
\address{\S\ Institute Instrumentation Center, IIT, Roorkee 247667,
  India}


\eads{\mailto{bhaduri@physics.mcmaster.ca}, 
\mailto{vandijk@physics.mcmaster.ca}}

\begin{abstract}
Quantum effects in statistical mechanics are important 
when the thermal wavelength is of the order of, or greater
than, the mean interatomic spacing. This is examined at depth taking
the example of a hard-sphere Boltzmann gas. Using the virial expansion
for the equation of state, it is shown that the interaction energy of
a classical hard-sphere gas is exactly zero. When the (second) virial
coefficient of such a gas is obtained quantum mechanically, however, 
the quantum contribution to the interaction energy is shown to be 
substantial. The importance of the semiclassical corrections to the
interaction energy shows up dramatically in such a system.
\end{abstract}

\submitto{EJP}


\maketitle

\section{Introduction}

The virial expansion of a dilute gas is standard fare in the
curriculum of the statistical mechanics course for beginning graduate
students.  The hard-sphere classical gas provides a nice example on
this topic, since the first few classical virial coefficients are
analytically known.  The quantum corrections in the dilute hard-sphere
gas, however, is a topic not covered in text books.

The objective of this paper is to study the quantum corrections in a
hard-sphere one-component Boltzmann gas. The usefulness of
semiclassical approximations in understanding the difference between
the classical and quantum results is demonstrated in this paper.  As
usual, we consider the gas in the thermodynamic limit, where the
number of atoms (of one kind only), $N\rightarrow\infty$, as also the
volume of the enclosure, $V\rightarrow\infty$, but the number density
$n=N/V$ remains finite. Each atom is taken to be a rigid
(impenetrable) sphere of diameter $a$, so that the minimum relative
distance between the spheres is $a$. It must be mentioned at the
outset that this system has been studied extensively in the last fifty
years (see \cite{hecht98} for a nice summary). Semiclassical quantum
corrections in the hard-sphere Boltzmann gas were examined (amongst
others) by Hill~\cite{hill68} and Gibson~\cite{gibson70}. Our focus in
this paper will be mainly on the interaction energy of the hard-sphere
gas. We first show, using the virial expansion for the equation of
state, that this is exactly zero when the virial coefficients for the
hard-sphere gas are calculated classically. {\it Since we consider a
  Boltzmann gas, there are no exchange contributions due to
  statistics}. For a dilute gas at high temperature, it is the second
virial coefficient that gives the dominant contribution. The second
virial coefficient may also be calculated quantum mechanically from
the Beth-Uhlenbeck formula using the hard-sphere scattering phase
shifts in the various partial waves~\cite{beth37}.  Contrary to naive
expectations, the quantum result for the interaction energy of the
Boltzmann gas at high temperatures is substantially different from the
classical one, even when the thermal wavelength is only a fraction of
the hard-sphere diameter.  Obtaining the quantum result requires
careful numerical work, adding contributions from more and more
partial waves with increasing temperature. The difference between the
quantum and the classical results may be accounted for by the
semiclassical corrections, given in the form of an asymptotic series.
This example shows the usefulness of the semiclassical method in
statistical mechanics even for a singular interaction.

\section{The interaction energy}

\subsection{The virial expansion}

To appreciate this in more detail, note that the virial expansion of 
a (one-component) gas (classical as well as quantal) for the pressure $P$ at a
temperature $T$ is given by~\cite{pathria72} 
\beq 
\frac{P}{n\tau}=1 + a_2(n\lambda^3) + a_3(n\lambda^3)^2 +\cdots, 
\label{state}
\eeq where $a_2, a_3,\dots,$ are the dimensionless second, third,
etc., virial coefficients, $\tau=k_BT$, $n=N/V$ is the number density
of particles, and $\lambda=\sqrt{2\pi\hbar^2\!/m\tau}$ is the thermal
wavelength.  Even for an ideal quantum gas of bosons or fermions, the
virial coefficients are nonzero. These contributions due to statistics
will not be included in our Boltzmann gas; only the part due to the
interaction between the atoms will be considered.  The free energy
$F=E-TS$ may be easily obtained by integrating the above $P$ with
respect to the volume $V$ (since $P=-(\partial F/\partial V)_{\tau}$),
the constant of integration being chosen to match the perfect
Boltzmann gas value:
\begin{equation}
F  =  -N\tau \ln \left(\frac{V}{\lambda^3}\right) + a_2 \lambda^3\tau
\frac{N^2}{V}+ \cdots 
  + a_j\lambda^{3(j-1)}\tau \frac{N^j}{(j-1)V^{j-1}}+\cdots.
\end{equation} 
The energy may next be obtained by using the relation 
$E=-\tau^2[\partial (F/\tau)/\partial\tau]$. After
subtracting out the energy of the perfect gas part, one obtains the 
virial series for the interaction energy per particle:  
\begin{eqnarray}
\frac{E_\textrm{int}}{N} & = & 
\frac{3}{2}\tau\left\{(n\lambda^3)\left[a_2-\frac{2}{3}
\tau\frac{da_2}{d\tau}\right]+(n\lambda^3)^2\left[a_3-\frac{2}{3}
\tau\frac{da_3}{d\tau}\frac{1}{2}\right]+\cdots \right. \nonumber \\
&& \left. +(n\lambda^3)^{(j-1)}\left[a_j-\frac{2}{3}
\tau\frac{da_j}{d\tau}\frac{1}{j-1}\right]+\cdots\right\}
\label{interaction}
\end{eqnarray}
The above equation for the interaction energy, modified for a
two-component Fermi gas, has been used recently by Ho and Mueller~\cite{ho04}. 
Whereas for a classical gas the virial coefficients may be
expressed as integrals involving the interaction potential, for the
quantum problem a solution of the $j$-body problem is needed to obtain
$a_j$. Henceforth we shall denote the classical virial coefficients by 
$\aa_j$ to differentiate from their quantum counter parts,  
to be still denoted as $a_j$.  

It is easy to see now why the interaction energy for a classical
hard-sphere gas is zero order by order, in every power of
$(n\lambda^3)$.  For a hard-sphere diameter $a$, ${\cal
  A}_2=\frac{2\pi}{3}(a/\lambda)^3$ (see \eref{class} below), and
the higher order ${\cal A}_j$'s may be expressed as ${\cal A}_j=\kappa
({\cal A}_2)^{j-1}$, where $\kappa$ is a constant~\cite{hecht98}.
Substituting this in \eref{interaction}, we see that the
interaction energy of the classical hard-sphere gas vanishes
identically in each power of $(n\lambda^3)$. In fact, the classical
hard-sphere gas is rather special, since the right hand side of
\eref{state} is independent of temperature.

We note that in the regime $(n\lambda^3)<\!< 1$,  the series 
(\ref{interaction}) may be terminated after the first order term  
\beq
\frac{E_\textrm{int}}{N}=\frac{3}{2}\tau\left\{(n\lambda^3)\left[a_2-\frac{2}{3}
\tau\frac{da_2}{d\tau}\right]\right\}~.
\label{term}
\eeq

The classical second virial coefficient $\aa_2$ is given by~\cite{pathria72} 
\beq
\aa_2(\beta)=\frac{2\pi}{\lambda^3}\int_0^{\infty}dr~ r^2 
\left(1-\exp(-\beta V(r))\right)~,
\label{class}
\eeq
where $\beta=1/\tau$, and $V(r)$ is the interatomic two-body potential.
The quantum second virial
coefficient requires a knowledge of the bound-state and the continuum 
spectra. For a one-component quantum gas, $b_2=-a_2$ is given 
by~\cite{pathria72} 
\bea
b_2(\beta)&=&\sum_{l=0}^{\infty}(2l+1)b_2^{(l)}~\nonumber \\
b_2^{(l)}(\beta)&=& \sqrt{2}\left[\sum_n \exp(-E_{n,l}\beta)
  +\frac{1}{\pi}\int_0^{\infty}dk~\frac{\partial\delta_l}{\partial k} 
\exp(-\frac{\hbar^2 k^2\beta}{m})\right]~.
\label{delta}
\eea In the above, $E_{n,l}$'s are the binding energies of the two-body
bound states (absent in our problem), and $\delta_l$ is the scattering
phase shift in the $l$th partial wave.  It is important to note that
our \eref{delta} differs from the usual definition given for
quantum gases in text books~\cite{pathria72,huang65}: the overall
multiplicative factor in our case being $\sqrt{2}$ rather than $2
\sqrt{2}$. The second virial coefficient, after all, is just the
two-body partition function for relative motion, and for the Boltzmann
case it has a factor $1/N!$, which explains why our factor is less by
$2$ for the two-body case. Our definition here is in agreement with
the earlier authors~\cite{hill68,gibson70,dewitt62} who considered the
Boltzmann case.

\subsection{Numerical comparison}
We compare the classical second virial coefficient
$\aa_2=\frac{2\pi}{3}(a/\lambda)^3$ with its quantum counter part
$-b_2$ as defined by \eref{delta} for the hard-sphere gas. The
phase shift $\delta_l$ for hard-sphere scattering is given
by~\cite{flugge74} $\tan \delta_l=j_l(ka)/n_l(ka)$, and a little
algebra shows that 
\beq
\frac{d\delta_l}{dk}=-\frac{a}{(ka)^2}~\frac{1}{j_l^2(ka)+n_l^2(ka)}~.
\eeq 
For a fixed $a$, as one goes to higher temperatures (i.e., larger
values of $(a/\lambda))$, more and more $l$ values are needed to get
convergence in the computed $b_2^{(l)}$ using \eref{delta}. For
example, for $(a/\lambda)=2$, convergence is obtained only when the
sum extends to $l_\mathrm{max}=33$~\footnote[5]{The quantity $b_2$ was
  calculated by first summing over $l$ and then integrating.  We
  specified a minimum relative error in the integrand (or the sum) of
  10$^{-3}$.}.  This may be understood by noting that for small $
\beta$ (high temperature), the damping of the integrand from the
gaussian is less effective, and higher values of the variable $k$ are
needed for the convergence. From impact parameter
argument~\cite{messiah61}, the cut off $l_\mathrm{max}\simeq (ka)$ in
the sum over $l$ therefore also increases.
\begin{figure}[htpb]
  \centering
   \resizebox{5.0in}{!}{\includegraphics{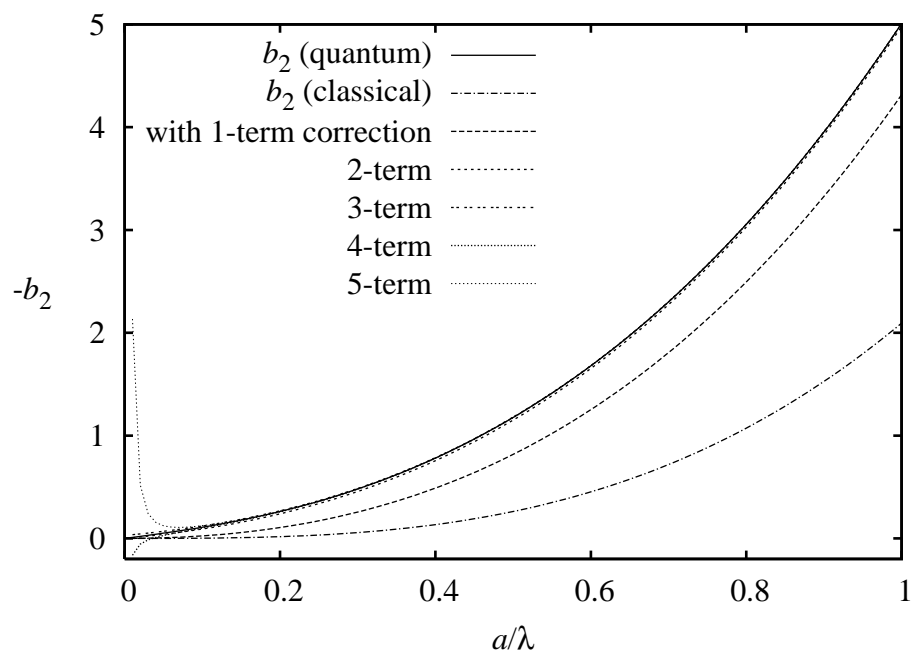}}
\caption{The quantum and classical second virial coefficient with 
  semiclassical corrections (to be discussed later.)  The lowest curve
  gives the classical second virial coefficient and semiclassical
  approximations approach the quantum mechanical curve for
  sufficiently large $\alpha/\lambda$.}
\label{figure_1}
\end{figure}
In \fref{figure_1}, we display the classical and quantum second
virial coefficients ($\aa_2$ and $-b_2$) as a function of
$(a/\lambda)$. One expects the classical regime for $(a/\lambda)>\!>1$.
As is clear from \fref{figure_1}, however, the two remain very
different even in the so called classical regime. This difference may
be understood from semiclassical corrections, as we shall see in the
next section. For a perfect Boltzmann gas, the energy per particle,
$E_0/N =\frac{3}{2} \tau$. 
\begin{figure}[htpb]
  \centering
   \resizebox{5.0in}{!}{\includegraphics{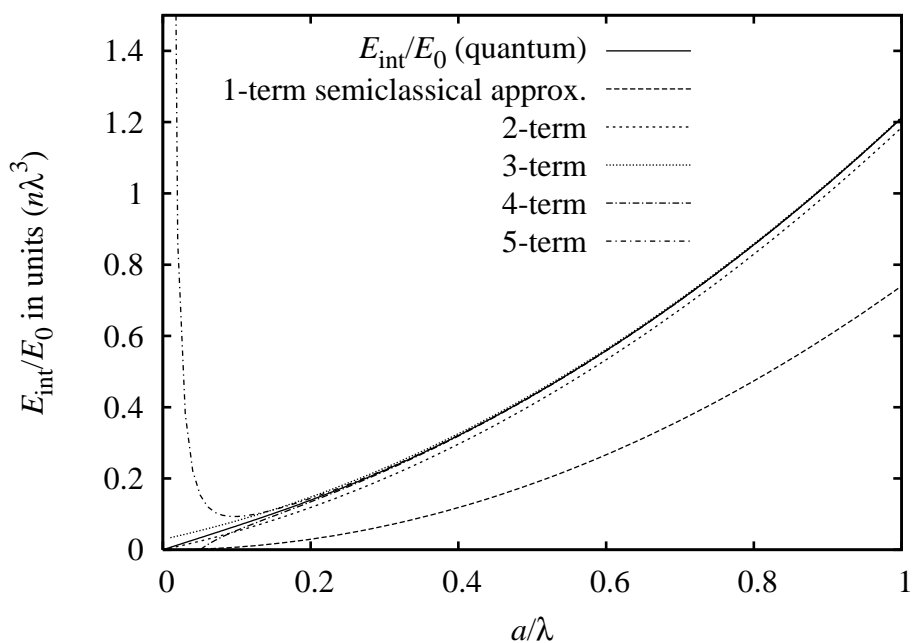}}
\caption{The quantum and semi-classical approximations to the 
  interaction energy in units of $n\lambda^3$.}
\label{figure_2}
\end{figure}
In \fref{figure_2}, we plot $E_\mathrm{int}/E_0$ in units of
$(n\lambda^3)$.  Recall that this this quantity is exactly zero
classically for any value of $(n\lambda^3)$. In \fref{figure_2},
the interaction energy is calculated from \eref{term} to first
order in $(n\lambda^3)$, using the virial coefficient $(-b_2)$ given
by \eref{delta}. The disagreement between the classical and the
quantum values of the second virial coefficient, as also the
interaction energy, keeps increasing for larger values of
$(a/\lambda)$. This, however, is misleading, as we see below. For
fixed $a$ and $n$, larger values of $(a/\lambda)$ imply smaller
$(n\lambda^3)$, which actually decreases the pressure $P/(n\tau)$ in
\eref{state} and the energy ratio $E_\mathrm{int}/E_0$ in magnitude.

To get a better feel for \fref{figure_2}, note that the maximum
density for close packing of the hard spheres~\cite{hecht98} is
$n_\mathrm{max}=\sqrt{2}/a^3$. In a dilute gas, let us
assume that $n=n_\mathrm{max}/10=\sqrt{2}/(10 a^3)$. We then
get 
\beq 
\left(\frac{a}{\lambda}\right)^3=\frac{\sqrt{2}}{10 (n\lambda^3)}~.
\eeq 
If we assume $(n\lambda^3)=0.1$, we get $(a/\lambda)^3=\sqrt{2}$,
or $a/\lambda\simeq 1$. From \fref{figure_2}, we see that
$E_\mathrm{int}/E_0\simeq 0.12$ in this case as opposed to zero if this was
calculated classically. Semiclassical corrections, to be discussed
later, are still important in this regime. On the other hand, for a
gas at equilibrium at room temperature and pressure, $n\lambda^3$ is
at least ten times less, so quantum corrections are negligible.

\begin{figure}[thpb]
  \centering
   \resizebox{5.0in}{!}{\includegraphics{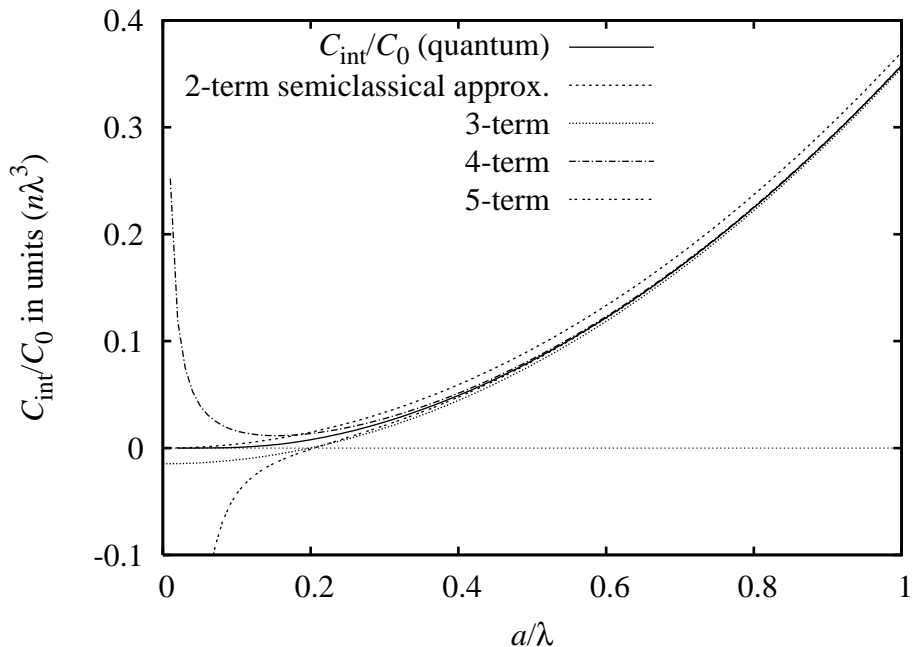}}
\caption{The specific heat calculated quantum mechanically and using 
  semi-classical approximations.  Since the second term in the
  semi-classical expansion is zero, the one-term and two-term
  approximations are identical.}
\label{figure_3}
\end{figure}

In \fref{figure_3}, 
we similarly plot $C_\mathrm{int}/C_0$ in units of $(n\lambda^3)$ as
a function of $(a\lambda)$. Classically. of course, 
$C_0$ is $3/2$ per particle in units of the Boltzmann constant, and 
$C_\mathrm{int}=0$.  

\section{The second virial coefficient}

\subsection{The classical-quantum connection}  

In this section, the connection between the classical $\aa_2(\beta)$
and its quantum counterpart $-b_2(\beta)$ will be explored for a given 
two-body potential $V(r)$. This will help us to understand why the
classical and quantum results are so different for the example that we
are studying. For this purpose, it is convenient to make the partial 
wave decomposition  
\beq
\aa_2(\beta)=\frac{2\pi}{3}\left(\frac{\lambda}{a}\right)^3=
\sum_{l=0}^{\infty}(2l+1) \aa_2^{(l)}~.
\eeq
it may be verified easily (by replacing the sum over $l$ by an
integral), that 
\beq
\aa_2^{(l)}=\frac{1}{\lambda}\int_0^{\infty}~ dr~ 
\exp\left[-\frac{\hbar^2l(l+1)}{mr^2}\beta\right][1-\exp(-\beta V(r))]~.
\label{partclass}
\eeq
One may enquire about the relation between this 
expression (\ref{partclass}) and the quantum counterpart
$b_2^{(l)}$. This question was actually answered nearly seventy years
back by Kahn~\cite{kahn38} in his doctoral dissertation. We shall give here a new
compact derivation of his result, considering only the part of the
virial coefficient arising from the scattering channel in the
continuum. Using
(\ref{delta}), and the relation $E=\hbar^2 k^2/m$, we have (with no 
bound states)    
\begin{equation}
b_2^{(l)}(\beta) =  \sqrt{2}
\left[\frac{1}{\pi}
\int_0^{\infty}dE~\frac{\partial\delta_l}{\partial E} 
\exp(-\beta E)\right]~.
\label{ho04}
\end{equation}
We note that $\frac{1}{\sqrt{2}} b_2^{(l)}(\beta)$ is just the
Laplace transform of the derivative of the phase shift with respect to
the energy. In the lowest order WKB approximation, the phase shift is
given by~\cite{messiah61}
\beq
[\delta_l(E)]_\textrm{WKB}=\sqrt{\frac{m}{\hbar^2}}\left[\left(\int_{r_l}^{\infty}
\sqrt{E-V_l(r)}-\int_{r_0}^{\infty}
\sqrt{E-\frac{\hbar^2 l(l+1)}{mr^2}}~\right) dr\right]~.
\label{wkb}
\eeq
In the above, the effective potential $V_l(r)$ is defined as 
\beq
V_l(r)=V(r)+\frac{\hbar^2l(l+1)}{mr^2}~.
\label{late}
\eeq 
In \eref{wkb}, $r_l$ and $r_0$ are the classical turning
points where the respective integrands go to zero. The derivative of
the phase shift may be written as 
\beq
\left[\frac{d\delta_l}{dE}\right]_\textrm{WKB}=\frac{1}{2}\sqrt{\frac{m}{\hbar^2}}\left[\left(
\int_0^{\infty}\frac{\Theta(E-V_l(r))}{\sqrt{E-V_l(r)}}-
\int_0^{\infty}\frac{\Theta(E-{\displaystyle \hbar^2l(l+1)\over 
\displaystyle {mr^2}})}
{\sqrt{E-{\displaystyle \hbar^2l(l+1)\over \displaystyle {mr^2}}}}\right)
dr\right]~
\label{wkb1}
\eeq
in terms of the unit step function $\Theta(x)$. By noting that the
Laplace transform of $\frac{\displaystyle \Theta(E-\mu)}
{\displaystyle \sqrt{E-\mu}}$ is 
$\sqrt{\frac{\displaystyle \pi}{\displaystyle \beta}}\exp(-\beta \mu)$, 
we immediately obtain the desired result
\begin{eqnarray}
\frac{1}{\sqrt{2}}[b_2]_\textrm{WKB}^{(l)} & = & 
-\frac{1}{\sqrt{2}\lambda} \int_0^{\infty}~dr
\exp\left[-\beta\frac{\hbar^2l(l+1)}{mr^2}\right]
[1-\exp(-\beta V(r))] \nonumber \\
& = & -\frac{\aa_2^{(l)}}{\sqrt{2}}~.
\label{nogo}
\end{eqnarray}
If the lowest order WKB approximation (\ref{wkb1}) for a given potential  
 were exact, the classical and quantum (partial wave)
virial coefficients would be the same.
In general, of course, there are higher order semiclassical 
corrections in powers of $\hbar$ to the WKB lowest order term that
accounts for the deviation from the classical result. We next examine
the semiclassical corrections.


\subsection{Semiclassical corrections to $\aa_2(\beta)$}


There is the Wigner-Kirkwood~\cite{wigner32,kirkwood33,brack03} 
semiclassical expansion of 
the quantum partition function (or the second virial coefficient) in powers 
of $\hbar^2$ and the derivatives of the potential. This, however, is
not applicable for the hard-sphere potential, since its derivatives do
not exist. For the hard-sphere problem Hill~\cite{hill68} and 
Gibson~\cite{gibson70} used advanced techniques in scattering theory to 
obtain the following asymptotic expansion for the virial coefficient :
\begin{eqnarray}
  b_2 & = & \aa_2\left[1+\frac{3}{2\sqrt{2}}\left(\frac{\lambda}{a}\right)
    +\frac{1}{\pi}
    \left(\frac{\lambda}{a}\right)^2+\frac{1}{16 \pi
      \sqrt{2}}\left(\frac{\lambda}{a}\right)^3 \right. \nonumber \\
 && \left.  - \frac{1}{105 \pi^2}
    \left(\frac{\lambda}{a}\right)^4+\frac{1}{640\pi^2 \sqrt{2}}
    \left(\frac{\lambda}{a}\right)^5+\Or\left(\left(
        \frac{\lambda}{a}\right)^6\right)\right],
\label{semi}
\end{eqnarray}
where the classical virial coefficient $\aa_2=\frac{2\pi}{3}
(a/\lambda)^3$. Since $\lambda$ is proportional to $\hbar$, the above
series is in powers of $\hbar$, and not $\hbar^2$. This is
characteristic of potentials with sharp surfaces~\cite{dewitt62}.  The
derivation of the above series is quite involved, and will not be
given here. We can present, however, a qualitative argument to
indicate why one may expect powers of $(\lambda/a)$ in the
semiclassical series above.  Consider the integral (\ref{class}) for
$\aa_2(\beta)$ for the hard spheres. The potential $V(r)$ is zero
whenever the relative distance $r$ between the centers of the spheres
is greater than $a$. For $r=a$ the spheres touch each other with their
sharp surfaces.  Semiclassically, though, we may expect a fuzziness
(or uncertainty) in the contact distance by an amount which is of the
order of the de Broglie wave length $\lambda_d=\hbar/\langle
p\rangle$, where $p$ is the absolute value of the relative momentum
between the two spheres. The average $\langle p\rangle$ with Boltzmann
distribution, and the corresponding de Broglie wavelength are given by
\beq 
\langle p \rangle =\frac{\displaystyle \int_0^{\infty} p^3
  \exp(-p^2\beta/m) dp} {\displaystyle \int_0^{\infty} p^2 \exp(
  -p^2\beta/m) dp} =\frac{2}{\sqrt{\pi}}\frac{m}{\beta},~~~
\lambda_d=\frac{\lambda}{2\sqrt{2}}~~.  
\eeq 
In evaluating the
integral (\ref{class}) semiclassically, we may put $V(r)=0$ for $r\geq
(a+\lambda_d)$, so that now the second virial coefficient becomes
$\frac{2\pi}{3}(a+\lambda_d)^3/\lambda^3$.  This may be written in the
form of a series in powers of $(\lambda/a)$, multiplied by the
classical term. This series, unlike the one given by \eref{semi},
terminates with the cubic term. Although the coefficient of the
leading $(\lambda/a)$ term is correctly reproduced, the agreement is
fortuitous. The main merit of this derivation is simply to show that
it is not unreasonable to expect semiclassical corrections in powers
of $(\lambda/a)$.

The order by order semiclassical corrections using \eref{semi} are
computed and shown for the second virial coefficient
(\fref{figure_1}), the ratio of $E_\mathrm{int}/E_0$
(\fref{figure_2}), and the specific heat ratio $C_\mathrm{int}/C_0$
(\fref{figure_3}). We see from these that the semiclassical
corrections are able to reproduce the quantum behaviour very well for
$(a/\lambda)>0.2$.  Since the series (\ref{semi}) is an asymptotic
one, the higher order terms are seen to diverge for small values of
$(a/\lambda)$.  Nonetheless, the series is very accurate for large
values of $(a/\lambda)$. For a given $(a/\lambda)$, the asymptotic
series under consideration has to be terminated after an optimum
number of terms are summed.  

We have thus demonstrated the usefulness of the semiclassical series
for understanding the classical to quantum behaviour.

\ack The authors (RKB \& WvD) would like to thank the Natural Sciences
and Engineering Council of Canada (NSERC) for the financial support of
this research.

\section*{References}


\providecommand{\newblock}{}

\end{document}